\pdfoutput=1
\documentclass[%
  prb,%
  letterpaper,%
  twocolumn,%
  showpacs,%
  floatfix,%
  superscriptaddress%
]{revtex4}
\RequirePackage{graphicx,amsmath,amssymb,bm}
\RequirePackage{hyperref}

\newcommand{\Fref}[1]{Fig.~\ref{#1}}

\newcommand{\Eqref}[1]{Eq.~(\ref{#1})}
\renewcommand{\eqref}[1]{eq.~(\ref{#1})}
\newcommand{\uvec}[1]{\boldsymbol{\hat{\textbf{#1}}}}
\def\lsim{\lower.35em\hbox{$\stackrel{\textstyle<}{\textstyle\sim}$}}
\def\gsim{\lower.35em\hbox{$\stackrel{\textstyle>}{\textstyle\sim}$}}

\usepackage[usenames,dvipsnames]{xcolor}
\usepackage{soul}

\def\lsim{\lower.35em\hbox{$\stackrel{\textstyle<}{\textstyle\sim}$}}
\def\gsim{\lower.35em\hbox{$\stackrel{\textstyle>}{\textstyle\sim}$}}

\begin{document}

\newcommand{\abs}[1]{\lvert#1\rvert}
\title{Emergent magnetic texture in driven twisted bilayer graphene}

\author{D. A. Bahamon}
\affiliation{MackGraphe\,--\,Graphene and Nano-Materials Research Center,
Mackenzie Presbyterian University, Rua da Consola\c{c}\~{a}o 896, 01302-907,
S\~{a}o Paulo, SP, Brazil}
\email{dario.bahamon@mackenzie.br}
\author{G. G\'omez-Santos}
\affiliation{
Departamento de F\'{\i}sica de la Materia Condensada, 
Instituto Nicol\'as Cabrera and Condensed Matter Physics Center (IFIMAC), Universidad Aut\'onoma de Madrid, E-28049 Madrid, Spain}
\author{T. Stauber}
\affiliation{Departamento de  Teor\'ia y Simulaci\'on de Materiales, Instituto  de Ciencias de Materiales de Madrid, CSIC, E-28049, Madrid, Spain}
\email{tobias.stauber@csic.es}

\date{\today}

\begin{abstract}
The transport properties of a twisted bilayer graphene barrier are investigated for various twist angles. Remarkably, for small twist angles around the magic angle $\theta_m \sim 1.05^{\circ}$, the local currents around the AA-stacked regions are strongly enhanced compared to the injected electron rate. Furthermore, the total and counterflow (magnetic) current patterns show high correlations in these regions, given rise to well-defined magnetic moments that form a magnetic Moir\'e superlattice. The orientation and magnitude of these magnetic moments changes as function of the gate voltage and possible implications for emergent spin-liquid behaviour are discussed.  
\end{abstract}

\maketitle

Despite its chemical simplicity, twisted bilayer 
graphene\cite{Lopes07,Suarez10,Bistritzer11,Moon12, 
PhysRevLett.108.216802,PhysRevB.93.035452} (TBG) hosts a number of surprising 
phenomena ranging from 
superconductivity,\cite{Cao18b,Yankowitz19,Moriyama19,Codecido19,Shen19,Lu19} 
correlated insulator phase,\cite{Cao18a,Choi19} emergence of a Hofstadter 
butterfly,\cite{Dean13,Kim17} anomalous Hall 
ferromagnetism,\cite{Sharpe19,bultinck2019,ZhangMao19} photonic crystal for 
nano-light,\cite{Sunku18} and intrinsic optical 
dichroism.\cite{Kim16,Zhang18,Zhang19,bultinck2019} There are also several new 
predictions such as chiral superconductivity,\cite{Lin19,Classen19} nematic 
phases,\cite{Kozii18} flat plasmonic bands,\cite{Stauber16} a longitudinal Hall 
effect\cite{Stauber18,Stauber18b}, long-lived plasmons\cite{Levitov19}, Moir\'e 
ordered current loops,\cite{Weckbecker19} and marginal Fermi 
liquid.\cite{Gonzalez19,Gonzalez19b} The tuneable twist angle thus changes the 
optical, plasmonic and electronic properties that may be used in novel {\it 
twisttronic }devices.\cite{Carr17} There are also related carbon systems such as 
twisted double-bilayer graphene\cite{Liu19,Cao19b,Shen19} or ABC-trilayer 
graphene on a $BN$-substrate\cite{Chen19} that show superconducting phases.  

The plethora of new phenomena is closely linked to the emergence of a new 
intermediate length scale given by the Moir\'e-lattice constant of $\sim$10 nm. 
This superstructure is due to the different crystallographic orientations of the 
two graphene layers and its periodicity can be defined by quasi-circular 
AA-stacked regions where the two graphene layers lie on top of each other. These 
islands are arranged in a triangular lattice, surrounded by AB- and BA-stacked 
graphene which are the dual configurations of Bernal-stacked bilayer graphene. 
At small twist angles and low energies, the wave-functions become 
quasi-localised\cite{Trambly10,Trambly12} within the AA-stacked regions, also 
leading to quasi-localised neutral collective modes.\cite{Stauber16} 

\begin{figure}[t]
\centering
\includegraphics[height=7cm]{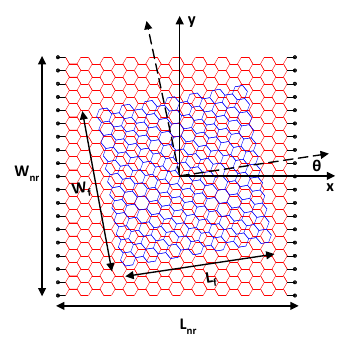}
\caption{ Schematic representation of the two-terminal system. The central part 
is a monolayer armchair graphene nanoribbon (red) of size $W_{nr} \times L_{nr} 
= 50~\text{nm}~\times~50~\text{nm}$ with a twisted graphene flake (blue) of size 
$W_{f} \times L_{f} = 40~\text{nm}~\times~40~\text{nm}$ on top. The black dots 
represent the effective contact region.}
\label{device} 
\end{figure}

The appearance of an intermediate Moir\'e scale should also influence the 
coherent transport properties, especially for filling factors around the charge 
neutrality point as it gives rise to the aforementioned flat-band physics. In 
this regime, a {\it paramagnetic} orbital response to an in-plane magnetic field 
is expected that becomes maximal around the magic angle 
$\theta_m=1.05^\circ$.\cite{Stauber18,Stauber18b} This response is caused by the 
so-called counterflow\cite{Bistritzer11,Liu17} where the current densities of 
the two layers move in opposite directions.\cite{Bistritzer11} 

In order to invoke this response electrically, an asymmetric driving with 
respect to the layers is required. Therefore, we shall here investigate the 
coherent transport properties of TBG by injecting electrons only in one of the 
layers. Our main result is the emergence of an AA-enhanced counterflow pattern, 
interpreted as a periodic magnetic texture due to the orbital motion of the 
electrons whose magnitude  can be tuned by the external source-drain voltage and 
whose triangular geometry might lead to frustration and spin-liquid behaviour.

\section{Model}

In order to investigate electronic transport through a TBG region with a large 
magnetic (or counterflow) response, we consider the system sketched in 
\Fref{device}. It consists of a monolayer armchair graphene nanoribbon of width 
$W_{nr}$ and length $L_{nr}$ with a graphene flake of size $W_f~\times~L_f$ on 
top. When the armchair edges of the nanoribbon and the patch are aligned we have 
an AB-stacked bilayer graphene region. Changing the orientation of the  flake, a 
TBG barrier is created for the electrons flowing in the nanoribbon. We choose 
the following dimensions: $W_{nr} = 50$ nm,  $L_{nr} = 50 $ nm,  $W_f = 40$ nm 
and $L_f = 40$ nm. This setup allows us to reduce the effect of corners, edges 
and evanescent states that would make the interpretation of the transport 
mechanisms more challenging. There are 97468 atoms in the bottom layer and 59432 
atoms in the top flake. The number of atoms connected to the source and drain 
contacts amounts to 237. 

Although in our setup we can impose  any twist angle, we will work with  
commensurate rotation angles $\cos(\theta) = 1-\frac{1}{2(3i^2 + 3i + 1)}$ to 
facilitate notation and comparison to the continuum model.\cite{Lopes07} The 
conductance is calculated within the Landauer-B\"uttiker formalism 
$G=\frac{2e^2}{h}\text{Tr}[\Gamma_LG_C\Gamma_RG_C^{\dagger}]$, where $G_C = 
[E-H_C-\Sigma_L-\Sigma_R]^{-1}$ is the Green's function of the central region 
and $\Sigma_L $, $\Sigma_R$ are the the self energies of the left and right 
contact, respectively; in the same way,  $\Gamma_{L} = i(\Sigma_L-\Sigma_L^{\dagger})$ 
and $\Gamma_{R} = i(\Sigma_R-\Sigma_R^{\dagger})$ define 
the coupling functions of the central region to the left and right contact.  

To describe the low energetic electronic properties, we use a tight-binding 
Hamiltonian (above denoted as $H_C$) with  hopping amplitude $t_{ij}(d_{ij}) =  
V_{pp\sigma}(d_{ij})\cos^2(\alpha) + V_{pp\pi}(d_{ij})\sin^2(\alpha)$ between 
sites $i$ and $j$, where $d_{ij} = |\vec{d}_{ij}| = |\vec{R}_j - \vec{R}_i|$ is 
the bond length and $\alpha$ is the angle formed by $\vec{d}_{ij}$ and the 
\textit{z-axis}. The value of the inter-atomic matrix elements is a function of 
the bond length:\cite{Brihuega12,Moon12} $V_{pp\sigma} = V_{pp\sigma}^0 
e^{-\frac{d_{ij}-d_0}{\delta}},~V_{pp\pi} = V_{pp\pi}^0 
e^{-\frac{d_{ij}-a}{\delta}}$ where $V_{pp\sigma}^0 = t_{\perp}^0 
=0.48~\text{eV}$,  $V_{pp\pi}^0 = t_0 =  -2.7~\text{eV}$, $a= 0.142~\text{nm}$, 
$d_{0} = 0.335~\text{nm}$ and $\delta = 0.184\sqrt{3}a$. 

To properly characterize the electronic properties of TBG, one needs to go 
beyond the nearest-neigbour description, thus for a site $i$ we select the 
neighbours $j$ located inside a radius $d_{ij}\leq 4a$. The previous restriction 
reduces the efficiency of recursive techniques to calculate the  Green's 
functions of the central region and the contacts. To overcome these technical 
problems, we first represent the Hamiltonian of the entire device as a sparse 
matrix and secondly, we set the self-energy terms to $\Sigma_{L}=\Sigma_{R}= 
-i|t_0|$.\cite{Bahamon} The last approximation is valid whenever a large number 
of modes is injected into the central region.\cite{Schomerus} 

\begin{center}
\begin{figure}[t]
\scalebox{0.85}{\includegraphics{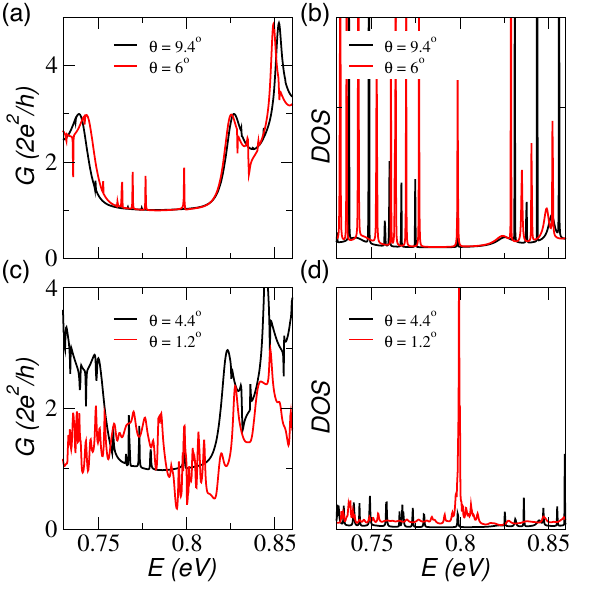}}
\caption{ Conductance (a) and DOS (b)  for $\theta = 9.4^{\circ}$(i=3) and 
$\theta = 6^{\circ}$(i=5).  Conductance (c) and DOS (d) for $\theta = 
4.4^{\circ}$(i=7) and $\theta = 1.2^{\circ}$(i=27).}
\label{tedos} 
\end{figure}
\end{center}

\section{Conductance}

There are three electronic regimes in TBG.\cite{Lopes07,PhysRevB.93.035452} For 
large rotation angles $\theta > 10^\circ$, the layers are basically decoupled at 
low energies. For intermediate angles with $2^\circ < \theta < 10^\circ$, the 
two layers become coupled leading to a renormalisation of the Fermi velocity and 
the emergence of van Hove singularities inside the first Moir\'e 
band.\cite{Luican11} For small angles $\theta < 2^\circ$, the layers are 
strongly coupled and low-energetic wave-functions become 
localized.\cite{Trambly10,Trambly12} 

We can observe these regimes also in the conductance of our system where we are  
especially interested in the transport properties around the charge neutrality 
point (CNP). Due to self-doping effects, the chemical potential of the CNP is no 
longer at zero energy, however, we can find it  gradually increasing the 
strength of the  TBG barrier.  For large angles $\theta > 10^\circ$, the TBG 
barrier is completely transparent and  the conductance of the single layer 
nanoribbon is not modified by the  twisted patch on top. Reducing the twisting 
angle, in the intermediate coupling regime for $\theta = 
9.4^\circ~\text{(i=3)},~6^\circ~\text{(i=5)},~4.4^\circ~\text{(i=7)}$, we see in 
 \Fref{tedos} resonant peaks superimposed on the first conductance plateau as 
well as the  reduction of the width of the  plateau. Interestingly,  the 
conductance and DOS peak at $E = 0.291t_0\sim0.8~\text{eV}$ appears for all 
twisting angles and allow us to pinpoint the CNP. To understand this, note that 
the contacts always inject electrons with a defined momentum $k_c$, while at the 
CNP  the TBG has only one transmission state with $k_x = 0$. This momentum 
mismatch\cite{chico} produces a quasilocalized state (see ESI\dag) and 
introduces an additional channel for transport at the CNP.  

In the strong coupling regime $\theta = 1.2^\circ$(i=27), the conductance 
quantization is completely gone and rapid oscillations around the CNP appear. In 
general, frequency as well as intensity of these oscillations become stronger as 
the twist angle is reduced. The physical origin behind these oscillations is the 
high Density of States (DOS) around the CNP, see \Fref{tedos} (d), i.e., the TBG 
barrier scatters the incident electrons into a large number of available states 
with the  same energy, causing the observed interferences. 

We also studied electronic transport for $\theta < 1^{\circ}$ (see ESI\dag),  
however, we do not observe high DOS around the CNP. For theses systems the 
Moir\'e periodicity \cite{Lopes07} $D = a/\sin(\theta/2) > 16.2~\text{nm}$, and  
there are few A-A stacked regions in our device to produce high DOS.
\section{Local current patterns.} 

It is important to  note that one atom in the bilayer region can have more than  
50 neighbors, thus the expression for the current between sites $i$ and $j$  

\begin{equation}
I_{ij} = \frac{2e}{h} \int_{E_F - eV_{SD}/2}^{E_F + eV_{SD}/2}   t_{ij} \left 
[G_{ji}^{<} - G_{ij}^{<} \right ]dE\;,
\label{eq:Iij}
\end{equation} 

\noindent where $G_{ij}^{<}$ is the lesser Green's function.\cite{Datta},  must 
be used cautiously to calculate the total current at one atomic site.  Running 
backwards from the magnetic moment definition based on bond currents, we  
provide a consistent definition of the total site current that, by construction, 
leads to the same magnetic moment. 

The classical image of a current-carrying {\em straight wire} from $\vec{r}_i $  
to $\vec{r}_j $ and the classical definition of magnetic moment, $ \vec m= 
\tfrac{1}{2}\int dV \vec{r}\times \vec{j}$, leads to the {\it global} definition 
of magnetic moment \cite{Walz}:

\begin{equation}\label{Mbond}
\vec m=\frac{1}{2}\sum_{<ij>}I_{ij}(\vec{r}_i\times \vec{r}_j)
,\end{equation}
where $\sum_{<ij>} $ means sum over pairs (bonds), counted once. The expression  
of \Eqref{Mbond} can be manipulated as follows:
\begin{equation}
\vec m=\frac{1}{2} \sum_i \sum_j  I_{ij} (\vec{r}_i\times  
(\vec{r}_j-\vec{r}_i))/2
,\end{equation} 
where we have used that $I_{ij}=-I_{ji} $, $I_{ij}(\vec{r}_i\times  \vec{r}_j) = 
I_{ji}(\vec{r}_j\times \vec{r}_i)$, $ \vec{r}_i\times \vec{r}_j = 
\vec{r}_i\times (\vec{r}_j-\vec{r}_i)$, and $\sum_{<ij>}=\tfrac{1}{2}\sum_i 
\sum_j $, where now $\sum_i $ (and $\sum_j $) runs over all sites. This leads to 
the following definition of site currents
\begin{equation}\label{Isite}
\vec{I}_i=\frac{1}{2a_{cc}}\sum_j I_{ij}(\vec{r}_j-\vec{r}_i)
,\end{equation}
and the associated expression for the magnetic moment
\begin{equation}\label{Msite}
\vec m=\frac{a_{cc}}{2}\sum_i \vec{r}_i\times \vec{I}_i
.\end{equation}
By construction, both expressions for the magnetic moment give the same answer,  
of course provided that local currents are associated to both sites of each 
bond.
Notice that, if one just wanted the same magnetic moment, the carbon-carbon  
distance $a_{cc}=0.142$ nm could be any number, provided it is the same object in 
\Eqref{Isite} and \Eqref{Msite}. The choice of a length for $a_{cc}$ is thus 
arbitrary and included for dimensional homogeneity of $I_{ij}$ and 
$\vec{I}_{i}$. \Eqref{Msite} can be interpreted as the discrete version of the 
textbook formula, $\vec m= \tfrac{1}{2}\int dV \vec{r}\times \vec{j} $, at least 
for a regular array of sites. 

\begin{center}
\begin{figure}[t]
\scalebox{1}{\includegraphics{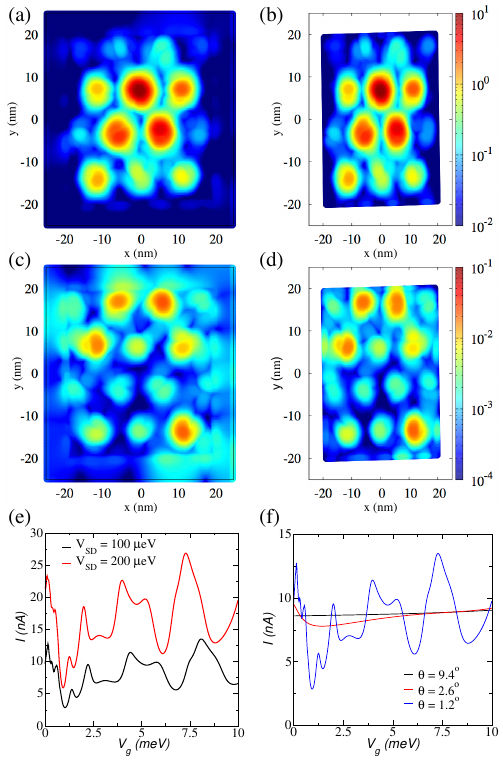}}
\caption{Magnitude of the electric current normalized by the total source-drain  
current per bond for  bottom and top layers:  $V_g \sim 0.1~\text{meV}$, panels 
(a) and (c), and $V_g \sim7~\text{meV}$, panels (b) and (d). Panel (e): 
Source-drain current as function of the gate voltage for two source-drain 
voltages $V_{SD} = 100,~200~\mu$eV. Panel (f): Source-drain current as function 
of the gate voltage for $\theta = 9.4^{\circ}(i=3), 
2.6^{\circ}(i=12)~\text{and}~1.2^{\circ}(i=27)$.  In all panels  $\theta = 
1.2^{\circ} (i=27)$ and $V_{SD} = 100~\mu \text{V}$ if no specified otherwise.} 
\label{Ireal} 
\end{figure}
\end{center}

\begin{figure*}
\centering
\scalebox{0.8}{\includegraphics{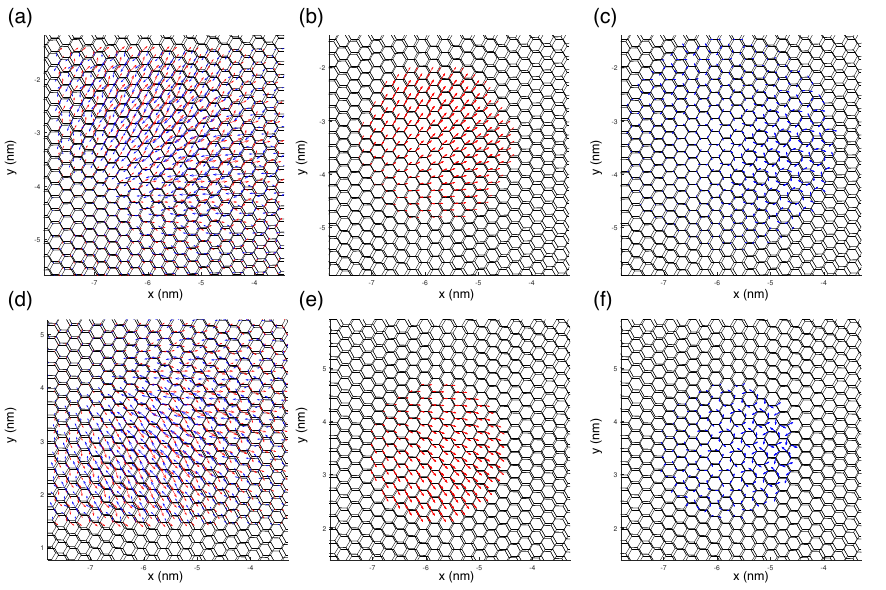}}
\caption{   For $\theta = 1.2^{\circ}$: (a) electric current, red arrow  
represents the electric current at atomic sites in the bottom layer 
($\vec{I}_1$) and the blue arrow for current in the top layer ($\vec{I}_2$). (b) 
counterflow(magnetic) current  and (c)  total current. For $\theta = 
-1.2^{\circ}$: (d) electric current in the bottom layer (red) and the top layer 
(blue).  (e) counterflow(magnetic) and (f)  total current per unit energy. In 
all panels $V_g \sim 0.1$ meV and $V_{SD}= 100~\mu\text{V}$.
}
\label{imap} 
\end{figure*}

We will now investigate the local current distribution through the TBG sample  
\cite{paez,Wakabayashi:2001yg,Liu:2004zr} using \eqref{Isite}.  Since,  we are 
interested in the transport properties around the CNP when high DOS is present,  
we select the twist angle $\theta = 1.2^{\circ}$. To make contact with 
experiment,\cite{Cao18b} we fix the bias voltage $V_{SD} = 100~\mu \text{V}$  
and study the current distribution as a function of  the gate voltage measured 
from the charge neutrality point ($V_g = E_F - 0.8~\text{eV}$). In \Fref{Ireal}, 
the magnitude of the current on  both layers is plotted for $V_g \approx 
0.1~\text{meV}$, see panels (a) and (b), and for $V_g \approx 7~\text{meV}$, see 
panels (c) and (d). The current is normalised by the average bond current 
injected into the drain contact $I_{SD}^{bond}$. For  $V_g=0.1~\text{meV}$, the 
total source-drain current $I \approx 12.5~\text{nA}$ is divided by the number 
of atoms connected to the drain contact, i.e., $I_{SD}^{bond} = 
12.5~\text{nA}/237$. 

It is clearly seen that there is a strong enhancement of the current in the  
AA-stacked regions which can be up to twenty times stronger compared to 
$I_{SD}^{bond}$ for $V_g=0.1~\text{meV}$. The formation of these hot spots can 
be linked to the enhanced density of states around the CNP observed on the 
AA-stacked regions. It can also be observed in the continuum model, see ESI\dag. 
Even more remarkable is the fact that the current intensity presents similar 
patterns and values on both layers  given the fact that the electrons are 
injected only into the bottom layer.   

To examine the origin  of the current in the top layer, we calculated the   
source-drain current  for large and small angles setting 
$V_{SD}=100~\mu\text{eV}$. The result is shown in \Fref{Ireal}(f). For $\theta = 
9.4^{\circ}$,  the current is carried by one transverse mode in  the bottom 
layer ($I_{SD}=\frac{2e}{h}V_{SD}$ where $\frac{2e}{h} = 80$ nA/meV). Around the 
CNP for  $\theta =2.6^{\circ}$, the source-drain current is slightly modified, 
however, when the current map is plotted we observe  similar magnitudes  in the 
top and bottom layer and hot spots on the AA-stacked regions (see ESI\dag). 
Given that  the total source-drain current is not reduced, we are forced to 
assume that the observed flow of charge is a response of the top layer to the 
injected current. 

\section{Chiral response and in-plane magnetic moment}

In order to develop a general analysis of the current response of the top 
layer, we highlight that an applied electric field induces an in-plane magnetic 
moment in chiral systems. To check if the size our device allows for a magnetic 
analysis we calculated current, DOS, LDOS and magnetic moment for $\theta = 
\pm1.2^{\circ}$. The infinite twisted bilayer system can be transformed from a 
positive to a negative twist angle by performing a parity-transformation 
$\vec{r}\to-\vec{r}$ and subsequent mirror-transformation ($\pi$ rotation around 
the $y$-axis). The position vector, current density, and magnetic moment 
transform accordingly, i.e., $(x,y,z)\to(x,-y,z)$, 
$(j_x,j_y,j_z)\to(j_x,-j_y,j_z)$, and $(m_x,m_y,m_z)\to(-m_x,m_y,-m_z)$.  We 
observe that LDOS, total magnetic moment and current in our finite system 
fulfill these requirements  for all energy points. Detailed results can be 
inspected in the ESI\dag.

For $V_g=0$ eV and $V_{SD} = 100~\mu\text{eV}$ in \Fref{imap}(a) and (d),  the 
current is shown in the bottom layer ($\vec{I}_1$) with a red vector and the 
current on the top patch ($\vec{I}_2$) by a blue vector for two mirror-symmetric 
 AA-stacked regions. The current pattern presents a complex structure, the 
number of neighbours considered in the tight-binding Hamiltonian and the 
interference effects introduce additional texture in the local current 
distribution. Despite of these circumstances, there exists a dominant current 
flow in all AA-stacked regions that produces an orbital magnetic moment. For 
this, we define the magnetic (or counterflow) and the total current:

\begin{equation}
\begin{split}
\vec{I}_{m} &= (\vec{I}_1-\vec{I}_2)/2 \\
\vec{I}_T &= \vec{I}_1+\vec{I}_2
\end{split}
\end{equation}

\noindent The magnetic current is well aligned and follows the transformation  
rules presented above, see \Fref{imap} (b) for $\theta = 1.2^{\circ}$ and in 
\Fref{imap} (e) for $\theta = -1.2^{\circ}$. Note that the defined pattern of 
the magnetic current means that the current in both layers flows in opposite 
directions producing an in-plane magnetic moment.  Although the  magnitude of 
total current is about $100$ times smaller than the magnetic one, in \Fref{imap} 
(c) and (f) $\vec{I}_T$  shows a circulating pattern.

\begin{center}
\begin{figure}[t]
\scalebox{0.3}{\includegraphics{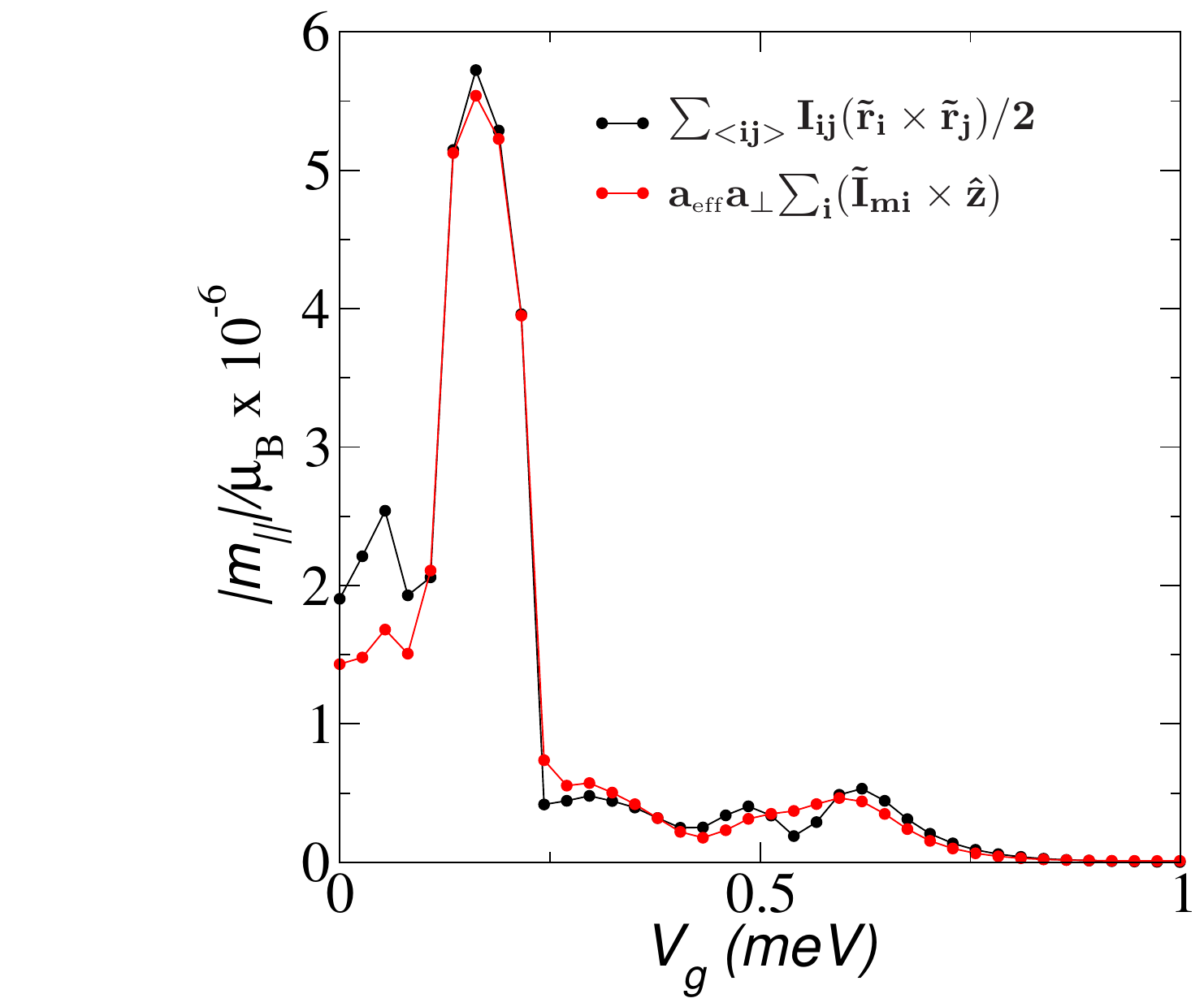}}
\caption{  In-plane magnetic moment per site as function of the gate voltage  
around the AA-stacked region between $-5\leq x/\text{nm} \leq 5$ and $2\leq 
y/\text{nm} \leq 12$.  The black line is calculated by the global definition of 
the magnetic moment ($\vec m=\frac{1}{2}\sum_{<ij>}I_{ij}(\vec{r}_i\times 
\vec{r}_j)$) while the red one by local definition ($\vec{m}_{\parallel} = 
a_{\text{eff}}a_{\perp}\vec{I}_{m} \times \uvec{z}$),  for $\theta = 1.2^{\circ} 
(i=27)$ and $V_{SD} = 100~\mu \text{V}$.}
\label{mmAA} 
\end{figure}
\end{center}

Having shown the simplicity of the magnetic current to describe the current  
response of the AA-stacked regions, let us now discuss the current response of 
the driven TBG sample in magnetic language. We can define the in-plane magnetic 
moment in two ways, i.e., globally and locally.
The first definition can be written in terms of the {\em bond}
current $I_{ij}$ as stated by \eqref{Mbond} for the whole device. The second  
definition involves the local magnetic current in the AA-stacked regions and 
reads

\begin{equation}
 \vec{m}_{\parallel} = a_{\text{eff}}a_{\perp}\vec{I}_{m} \times \uvec{z}\;,
 \label{eq:mmlocal}
\end{equation}
where $a_{\perp} = 0.335~\text{nm}$ is the interlayer distance and $\uvec{z}$  
the out-of-plane unit vector. We also introduced an effective bond length 
$a_{\text{eff}}$. Its value can be estimated projecting the bond length onto the 
vectors of the hexagonal lattice $a_{\text{eff}} = 
2(a+\frac{a}{2}+\frac{a}{2})/6\approx 0.67a$. For a  nearly prefect match 
between both approaches we set $a_{\text{eff}} = 0.7a$ (see \Fref{mmAA}). 

In \Fref{mmAA} we calculate the in-plane magnetic moment for the AA-stacked  
region between $-5\leq x/\text{nm} \leq 5$ and $2\leq y/\text{nm} \leq 12$ as 
function of the gate voltage. Clearly both approaches yield very similar results 
for the in-plane magnetic moment per site in units of Bohr magneton $\mu_B$. 
Although we present the results for one region,  other AA-stacked regions 
present similar behaviour. This is, non-zero in-plane magnetic moment around the 
CNP and oscillations. 

\begin{figure}[t]
\centering
\scalebox{0.81}{\includegraphics{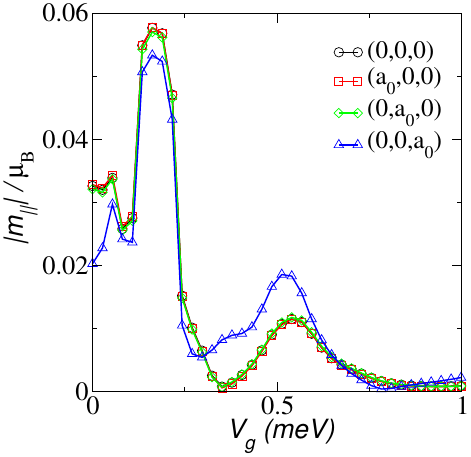}}
\caption{ Magnitude of the in-plane magnetic moment  for the whole device  
shifting the origin by  $\vec{R}_0=(0,0,0)$, $\vec{R}_0=(a_0,0,0)$, 
$\vec{R}_0=(0,a_0,0)$, and $\vec{R}_0=(0,0,a_0)$ where $a_0=10$ nm. The 
components of the magnetic moments are calculated using  $\vec 
m=\frac{1}{2}\sum_{<ij>}I_{ij}(\vec{r}_i\times \vec{r}_j)$ for a device with  
$\theta = 1.2^{\circ} (i=27)$ and $V_{SD} = 100~\mu \text{V}$.}
\label{mmshift} 
\end{figure}

It is important to mention that the expressions for the global and local  
magnetic moment are only well-defined if the total current $\vec{I}_{T} = 
\vec{I}_1+\vec{I}_2$ vanishes, i.e., in a closed system.\cite{PhysRevB.51.11584} 
This is not the case in our driven system as a net current flows through it, and 
any change of origin $\vec{r}_i\to \vec{r}_i-\vec{R}_0$ would lead to an 
additional contribution to the magnetic moment: $\vec{m}_{\vec{R}_0}= 
\frac{1}{2}\vec{R}_0\times \vec{I}_V$, with $\vec{I}_V=\sum_{<ij>}\vec{I}_{ij}$. 
 To find out under what circumstances the magnetic moment is well-defined, we 
shift the center of our infinite device (TBG barrier + contacts) to  positions 
$(a_0,0,0)$, $(0,a_0,0)$ and $(0,0,a_0)$ where $a_0=10$ nm and compare with the 
original system centred at  $(0,0,0)$. It is noticed in \Fref{mmshift} that the 
in-plane magnetic moment ($m_x$ and $m_y$) is well defined for low gate 
voltages. This is  nicely illustrated for  shifts constrained to the 
\textit{xy}-plane, where the perfect agreement is a consequence of $I_z = 0$. 
The differences observed for out-of-plane displacement can be traced back to a 
small, but non-zero $I_y$. 

\begin{center}
\begin{figure}[t]
\scalebox{1.0}{\includegraphics{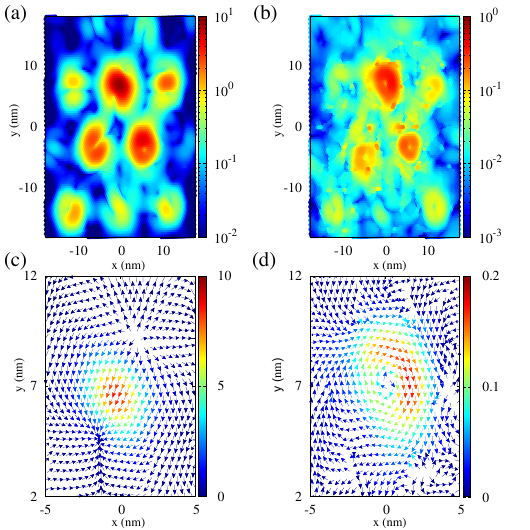}}
\caption{(a) Magnitude of the magnetic current or counterflow $\vec{I}_{m} = 
(\vec{I}_1-\vec{I}_2)/2$. (b) Magnitude of the total current $\vec{I}_{T} = 
\vec{I}_1+\vec{I}_2$. Vectorial map of the counterflow current (c) and total 
current (d) over one AA-Stacked region. In all panels the current is normalized 
by the total source-drain current per bond, $\theta = 
1.2^{\circ}(i=27)$,$V_g\sim 0.1~\text{meV}$ and $V_{SD} = 100~\mu \text{V}$. 
}
\label{Imagtot} 
\end{figure}
\end{center}

Still, the magnetic moment as obtained from the first definition is almost  
independent of the choice of the origin. This remarkable result can be 
understood by looking at the current map of the magnetic and total current whose 
absolute values differ by two orders of magnitudes, see \Fref{Imagtot} (a) and 
(b). This is particularly clear in the AA-stacked regions: zooming in on one of 
these regions, a strongly enhanced and highly oriented counterflow is 
appreciated, see \Fref{Imagtot} (c). Therefore, a well-defined local magnetic 
moment can be attached to each  AA-patch, because $\vec{I}_{m}\gg\vec{I}_{T}$. 
The presence of well-localised counterflow patterns must be accompanied by a 
source and drain. This can be seen from the vectorial map of the counterflow 
that clearly shows the presence of a source and a sink of the magnetic current 
on the AA-stacked region. The non-zero divergence of the magnetic current leads 
to a accumulation of charge on the two layers with opposite sign provoking a 
current in $z$-direction and thus closing the loop current. Or, to put it in 
another words, the total out-of-plane current $I_z = 0$  as  previously 
anticipated from the current response of the top layer and the shift of the 
origin. Let us also note that the direction of the angular magnetic moment of 
the several patches is not inferred by the source-drain direction, but that it 
is virtually random and tuneable by the gate-voltage within our finite sample. 
 
It is worth noticing that large orbital magnetic moments also apears in $C_{60}$ 
 molecular bridges\cite{PhysRevLett.87.126801} and carbon 
nanotubes,\cite{PhysRevB.75.153406} however, its appearance highly depends on 
the source and drain electrodes. This is not so in our case since our electrodes 
are not directly attached to the TBG barrier, in fact, they are far form the 
barrier by approx. 5 nm. Furthermore, linear response within the continuum model 
for bottom layer driven (infinite) TBG yields the same picture for the in-plane 
magnetic moment, i.e., large magnetic and small total currents in the AA-stacked 
regions. This is true for the whole band (not only around the neutrality point) 
and  it also confirms the peculiar nature of the 
excitations,\cite{Stauber18,Stauber18b} i.e., the current on the bottom layer is 
opposite to the applied source-drain voltage, see ESI\dag.

\subsection{Out-of-plane magnetic moment.} 

The general (global) definition also yields a finite magnetic moment 
in $z$-direction. This is consistent with a local interpretation since even 
though $|\vec{I}_T|$ is much smaller than $|\vec{I}_{m}|$, it is finite and 
shows a vortex structure on the scale of the AA-stacked region as seen in 
\Fref{Imagtot} (d). Moreover, at the atomic level we observe microscopic loop 
currents around the hexagonal plaquettes of single-layer graphene, giving rise 
to additional out-of-plane moments, see \Fref{imap}(c) and (f). Still, this 
effect seems not as robust as the counterflow.

\section{Robustness under perturbations} 

\begin{center}
\begin{figure}[t]
\scalebox{1.0}{\includegraphics{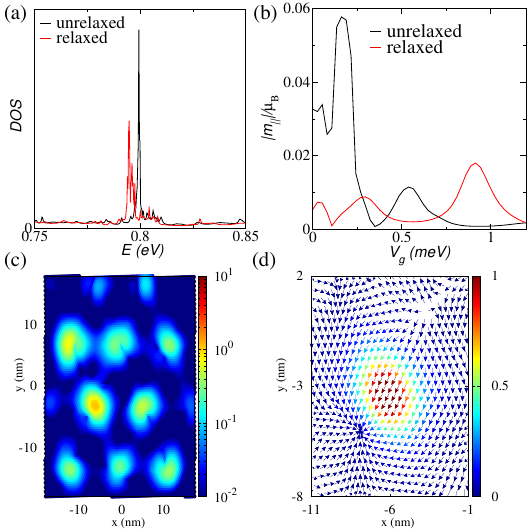}}
\caption{Results for the relaxed lattice with $\theta = 1.2^{\circ}(i=27)$: (a) 
DOS. (b) Magnitude of the in-plane magnetic moment  for the whole device 
calculated  by the global 
expression  $\vec m=\frac{1}{2}\sum_{<ij>}I_{ij}(\vec{r}_i\times \vec{r}_j)$. 
(c) Magnitude of the magnetic current or counterflow $\vec{I}_{m} = 
(\vec{I}_1-\vec{I}_2)/2$. (d) Vectorial map of the counterflow current over one 
AA-Stacked region. In all panels the current is normalized by the total 
source-drain current per bond, $V_g\sim 0.1~\text{meV}$ and $V_{SD} = 100~\mu 
\text{V}$. 
}
\label{fig:relax} 
\end{figure}
\end{center}

We will now analyze our results in the presence of various perturbations such 
as lattice relaxation, edge orientation, or edge disorder due to vacancies. It 
will turn out that the basic features such as the emergence of a magnetic 
texture are unchanged and should thus be experimentally observable.   

\subsection{Lattice relaxation}
The calculations presented so far assume that the individual graphene 
layers preserve their crystallographic structure when stacked on top of each 
other. However, differences among the binding energies of the AA and AB/BA 
stacked regions lead to lattice relaxation, and this  process reduces the area 
of the AA stacked region.  Thus, to address the robustness of the observed 
magnetic response we follow the procedure presented by Nam and 
Koshino to include in-plane lattice relaxation.\cite{Nam17,NamErratum} Although 
 the CNP is downshifted to $E \sim 0.79$eV, the relaxed lattice continues to 
present 
a high DOS around the CNP as represented by the red line in \Fref{fig:relax} (a). 
Importantly, all previously obtained features, i.e., high LDOS, high current and 
localized magnetic moments on the AA-stacked regions, are robust against lattice 
relaxation. However, if we compare the LDOS or the current for specific  
unrelaxed-relaxed lattice regions at the same $V_{SD}$ and $V_g$, we observe 
differences with respect to the actual numbers and local distributions. Thus, we 
cannot expect that the calculated magnetic moments for unrelaxed/relaxed regions 
present similar lineshapes as function of, e.g., the gate voltage, see 
\Fref{fig:relax} (b). Nevertheless, also in the relaxed lattice we find the 
emergence of a well defined magnetic texture as shown in \Fref{fig:relax} (c)-(d). 

Note that our results should thus be a general feature because the most  
important ingredient  to 
produce the magnetic texture is the wave function localization at the AA-stacked 
regions, and this effect is robust against in-plane\cite{Nam17,NamErratum} and 
out-plane lattice relaxation.\cite{PhysRevB.99.195419,Gargiulo_2017,Jain_2016}

\begin{center}
\begin{figure}[t]
\scalebox{1.0}{\includegraphics{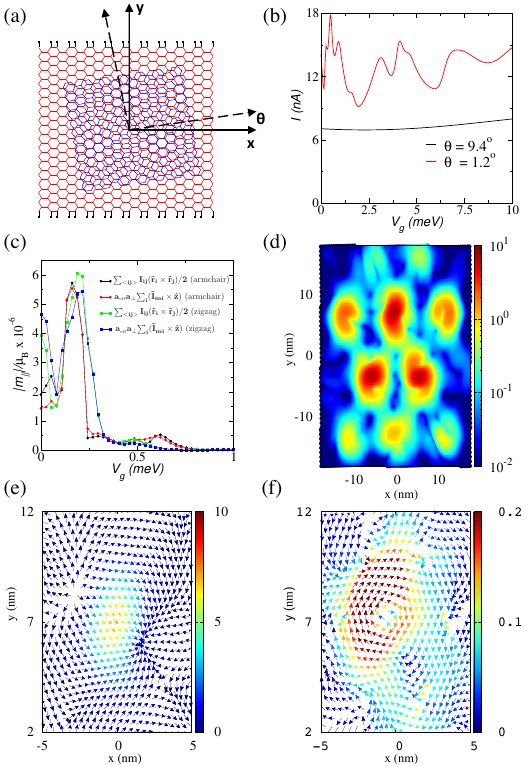}}
\caption{All panels present results for the TBG barrier embedded in a 
zigzag graphene nanoribbon. (a) Schematic representation of the device with 
zigzag edges, the central region and the twisted flake have  the same dimensions 
of the TBG on top of the armchair graphene nanoribbon ($W_{nr} \times L_{nr} = 
50~\text{nm}~\times~50~\text{nm}$ and $W_{f} \times L_{f} = 
40~\text{nm}~\times~40~\text{nm}$). The black dots represent the effective 
contact region, now at the top and bottom edge of the central region. (b)  
Source-drain current as function of the gate voltage for $\theta = 
9.4^{\circ}(i=3)~\text{and}~1.2^{\circ}(i=27)$. (c) In-plane magnetic moment per 
site as function of the gate voltage around the AA-stacked region between 
$-5\leq x/\text{nm} \leq 5$ and $2\leq y/\text{nm} \leq 12$. (d) Magnitude of 
the magnetic current or counterflow $\vec{I}_{m} = (\vec{I}_1-\vec{I}_2)/2$. 
Vectorial map of the counterflow current (e) and total current (f) over one 
AA-Stacked region. The current is normalised by the total 
source-drain current per bond. If not specified $\theta = 1.2^{\circ}(i=27)$,
$V_g\sim  0.1~\text{meV}$ and $V_{SD} = 100~\mu \text{V}$. 
}
\label{fig:zz} 
\end{figure}
\end{center}

\subsection{Zigzag edges}
With the aim of simplifying  the  initial analysis, we selected an 
armchair nanoribbon as the bottom monolayer of our device. However, it is well 
known that armchair and zigzag graphene nanoribbon present different electronic  
transport features. One of these properties is the spatial profile of the  
current that can have a significant impact on our result, e.g., for low 
energies, the 
current in  armchair graphene nanoribbons spreads uniformly over the width, 
while it is highly peaked at the center of the zigzag graphene nanoribbons. 
\cite{Zarbo2007} 

Now that we have a clear picture of the physical processes 
occurring within the TBG barrier under the asymmetric driving, we can focus  on 
the response of the TBG barrier embedded in a zigzag graphene nanoribbon (see 
\Fref{fig:zz} (a)). 
In \Fref{fig:zz} (b)-(f), we observe that there is not a qualitative 
difference between the response of the TBG barrier on top of an armchair or zigzag graphene nanoribbons, i.e., we continue to observe a strong enhancement of the current around the 
CNP and for low angles and strong in-plane 
magnetic moments on the AA-stacked regions. 

Let us analyse our results in more detail. First, it is worth mentioning that 
the magnitude of the in-plane magnetic moment for the AA-stacked region between 
$-5\leq x/\text{nm} \leq 5$ and $2\leq y/\text{nm} \leq 12$  presents similar 
values to those of the armchair case as shown in \Fref{fig:zz} (c). 
Second, the magnetic current of the  
underlying zigzag nanoribbon is rotated by $90^\circ$ compared to armchair case, 
shown in  \Fref{fig:zz}(e) and   
\Fref{Imagtot}(c), respectively. However, the total current preserves its  
vortex structure, plotted in 
\Fref{fig:zz}(f). 

\subsection{Edge disorder}
Let us finally address the robustness of the in-plane magnetic 
moments against edge disorder due to vacancies. Although we find a general  
reduction of the total magnetic moment in the presence of 10\% of vacancies in the 
the top flake edges, our 
main conclusions still remain unaltered. Details on the calculations can be 
found in the ESI.

\section{Conclusions.} 
We find a non-trivial texture of angular
orbital momentum which is necessarily arranged
in a triangular lattice. This is highly reminiscent of a Skyrmion lattice 
where the spin-texture is defined by circular domain walls which are arranged 
in a lattice configuration and recently seen in single layer 
graphene.\cite{zhou2019skyrmion} We speculate that similar physics might arise 
in our system. Here, the magnetic texture is highly tuneable since the induced 
magnetic moments are directly related to the source-drain voltage and can thus 
be changed from the quantum regime with small magnetic moments to a "more 
classical" regime with larger magnetic moments. More importantly, the magnetic 
moments in the AA-stacked regions are not polarized relative to the direction of 
the source-drain current and incommensurable twist angles should enhance this 
effect. The expected dipolar coupling between two localised magnetic moments 
might thus become important and eventually even lead to collective magnetic 
behaviour or even to a genuinely two-dimensional spin-liquid state.\cite{Han12} 
This should be measurable in transport or local probe experiments.

The dipolar interactions can directly be tuned by the twist angle which 
changes the lattice constant $L_M\sim\theta^{-1}$ of the triangular 
Moir\'e-lattice, i.e., for   $1^\circ\ \lsim\  \theta\ \lsim\ 2^\circ$  they would 
differ by a factor of $2^3\sim10$ assuming constant localised magnetic moments 
adjustable by the source-drain voltage. We also envision the possibility of 
electrical control of magnetic excitations, a long-sought goal in the quest of 
improved heat management for current technologies based on charged 
carriers.\cite{chumak2015magnon} Our setup might further be interesting in view 
of nano magnetism, usually concerned with using chiral and topological 
excitations such as skyrmions to store information in small volumes. Here, it 
would be the size of a Moir\'e unit cell.

In summary, we presented a transport study of a monolayer armchair and 
zigzag ribbons with a  twisted graphene flake on top. Around the neutrality 
point, the TBG barrier scatters electrons mainly into evanescent modes and for 
twist angles around the magic angle, the response is dominated by the large 
number of localized states on the AA-stacked regions. The high local DOS also 
gives rise to an enhanced localized counterflow when a source-drain voltage is 
applied to only one layer, resulting in a highly tuneable lattice of 
well-defined in-plane orbital magnetic moments with potential technological 
interest. 

In our finite sample, a high local DOS is only seen around the neutrality 
point, but we expect similar features to be observed for larger filling factors 
in macroscopic samples where the band-structure has fully developed. This is 
based on calculations made for the continuum model where the enhanced 
counterflow exists over the entire first conduction and valence band, 
respectively.  

\section*{Acknowledgements}
We acknowledge interesting discussions with Nuno Peres. This work has 
been supported by Spain's MINECO under Grant No. FIS2017-82260-P, 
PGC2018-096955-B-C42, and
CEX2018-000805-M as well as by the CSIC Research Platform on Quantum Technologies PTI-001. DAB acknowledges financial support from FAPESP (process nos. 2015/11779-4 and 2018/07276-5), CAPES  PrInt project no.  88887.310281/2018-00, CNPq process 306434/2018-0 and Mackpesquisa.

\begin{widetext}
{\Huge{\bf Supplementary Information}}

\begin{figure*}[h]
\centering
\scalebox{0.85}{\includegraphics{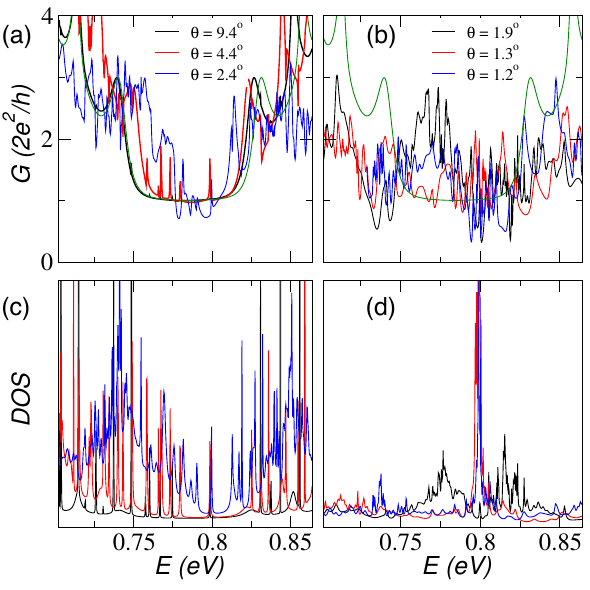}}
\caption{Conductance of the twisted bilayer graphene barrier in the (a) intermediate twist angle regime for $\theta = 9.4^{\circ}(i=3),~4.4^{\circ}(i=7)~\text{and}~2.4^{\circ}(i=13)$. (b) Small twist angle regime for $\theta = 1.9^{\circ}(i=17),~1.3^{\circ}(i=24)~\text{and}~1.2^{\circ}(i=27)$. The green line on both panels (a-b) corresponds to the conductance of monolayer armchair graphene nanoribbon. (c) Density of States (DOS) for twist angles presented in panel (a). (d) DOS for twist angles in panel (b).
}
\label{tedosSI} 
\end{figure*}

\subsection*{Conductance, momentum mismatch and charge neutrality point}

This section contains a detailed discussion of the conductance and Density of States (DOS) observed in  the twisted bilayer graphene (TBG) barrier. In \Fref{tedosSI} (a), the conductance for the TBG barrier is shown for intermediate twist angles. The strength of the TBG barrier for $\theta = 9.4^\circ (i=3)$ is still very weak and the conductance lineshape is similar to the conductance for twist angles $\theta > 10^{\circ}$ (green line in \Fref{tedosSI}a). For $\theta = 4.4^\circ (i=7)$, there are resonant peaks superimposed on the first conductance plateau. There is also a reduction on the width of the same plateau. For $\theta = 2.4^\circ (i=13)$ the first conductance plateau is strongly reduced.  

Irrespective of the rotation angle, there are two key attributes  in the conductance of the TBG barrier in  this twist angle regime: (i) the reduction of the twist angle introduces  a continuous set of conducting states, since there are energy regions with  conductance values higher than the ones obtained for the weak coupling regime. In those regions, interference is seen which is a consequence of having more than one conducting channel.\cite{gonzalez} (ii) There is a  conductance  peak at $E\sim 0.8~\text{eV}$. 

The physical origin of both attributes can be deduced from the DOS, shown in \Fref{tedosSI} (c).  The first conductance feature can be understood by noticing that new conducting states rise around $E\sim 0.8~\text{eV}$ when the twist angle is reduced. On the other hand, the transmission peak at $E\sim 0.8~\text{eV}$ signals the position of the charge neutrality point (CNP) of the TBG barrier. This can be understood by noticing that the highly doped left contact injects electrons with a defined momentum $k_c$ which are scattered into a number of available  states with momentum $k_x^{\text{TBGb}} =V_g/\hbar v_F$, where $V_g=E_F-E_{\text{CNP}}$ is the gate voltage\cite{Tworzydlo06} and $E_{\text{CNP}}$ the energy at the CNP. At the CNP, the  mismatch between the available momenta in the contacts ($k_c$) and  the TBG barrier ($k_x^{\text{TBG}} = 0$) produces an evanescent state and partial reflection at the barrier. The constructive interference between these waves produces a transmission peak.\cite{chico} 

Notice that the momentum mismatch and $E_{\text{CNP}}$ hardly depend on the twist angle and the peaks in the conductance and DOS observed for all twist angles at $E\sim 0.8$ eV  indicate the position of $E_{\text{CNP}}$. The appearance of the high DOS peak at the same energy in the small angle regime confirms that our transport analysis is correct. 

In the small angle regime ($0^\circ < \theta < 2^\circ$), the conductance quantization is completely gone. We can see in \Fref{tedosSI} (b) that the conductance shows rapid oscillations around $E\sim0.8\text{eV}$ and the frequency as well as the intensity of these oscillations increase as the angle is reduced. The TBG barrier thus again scatters incident electrons into different  channels with the same energy. However, in the small angle regime there is a higher DOS around $E\sim0.8\text{eV}$, shown in \Fref{tedosSI} (d). Consequently, the electrons transmit through a larger number of propagating states generating more complex interference patterns.  

\begin{figure*}[t]
\centering
\scalebox{0.9}{\includegraphics{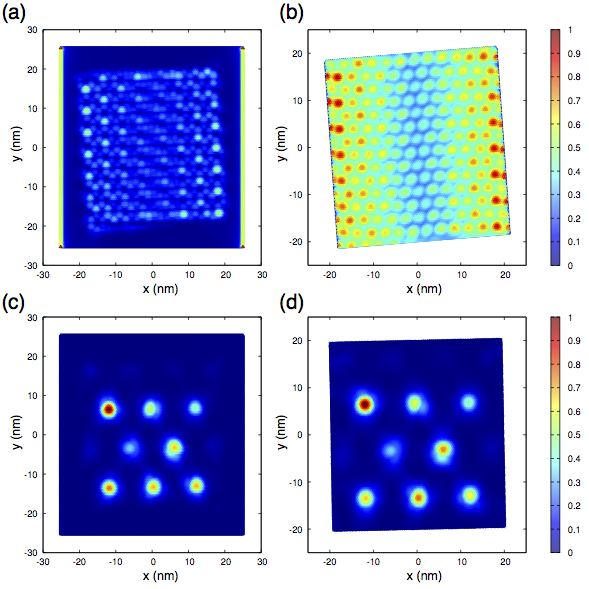}}
\caption{   Local Density of States for $\theta = 4.4^{\circ} (i=7)$ and $E= 0.8~\text{eV}$: (a) bottom layer; (b) top layer. Local Density of States for $\theta = 1.2^{\circ} (i=27)$ and $E_F= 0.8~\text{eV}$: (c) bottom layer; (d) top layer. 
}
\label{ldos} 
\end{figure*}

\subsection*{Local Density of States}

In  \Fref{ldos} (a) and (b),  we have plotted the local Density of States (LDOS) for the for the state at $E=0.8\text{eV}$ and twist angle $\theta = 4.4^{\circ}$. The state presents all the characteristics of an evanescent state: high LDOS at the edges that  decays towards the center of the TBG barrier. However, the top  and bottom layer are still weakly coupled since the LDOS is not evenly distributed over both layers. Consequently, in this regime the top patch behaves as an additional channel for the transport.

In the small angle regime, the interference discussed above is also appreciated looking at LDOS. In \Fref{ldos} (c) and (d), the LDOS is shown for $\theta = 1.2^{\circ}$ at $E=0.8\text{eV}$. The high LDOS is unevenly distributed over the  AA-stacked regions as a result of the multiple electronic paths. The lower LDOS in the regions close to the edges of the top graphene flake are finite size effects indicating a reduction of the  Moir\'e confinement potential. In the small angle regime, electrons thus transmit through the sample via a number of degenerate states localized on AA-stacked regions.

\subsection*{Charge neutrality of bulk TBG}

For an additional confirmation of the CNP location, we calculated the band structure for bulk TBG for different twist angles. The CNP of the bulk system is located around the same value we obtained from transport calculations of our finite system, see \Fref{bands}. 

Based on the above and the conductance calculation, we can confirm that our finite system reproduces the main features reported for bulk TBG for $\theta > 1^\circ$. Although DOS plots of our device show oscillations due to the confinement, we clearly observe: (i) new vHs with the reduction of the twist angle, (ii)  Merging of vHs for small angles and (iii) localization of the wave function on AA-stacked regions.

\begin{figure*}[h]
\centering
\scalebox{0.3}{\includegraphics{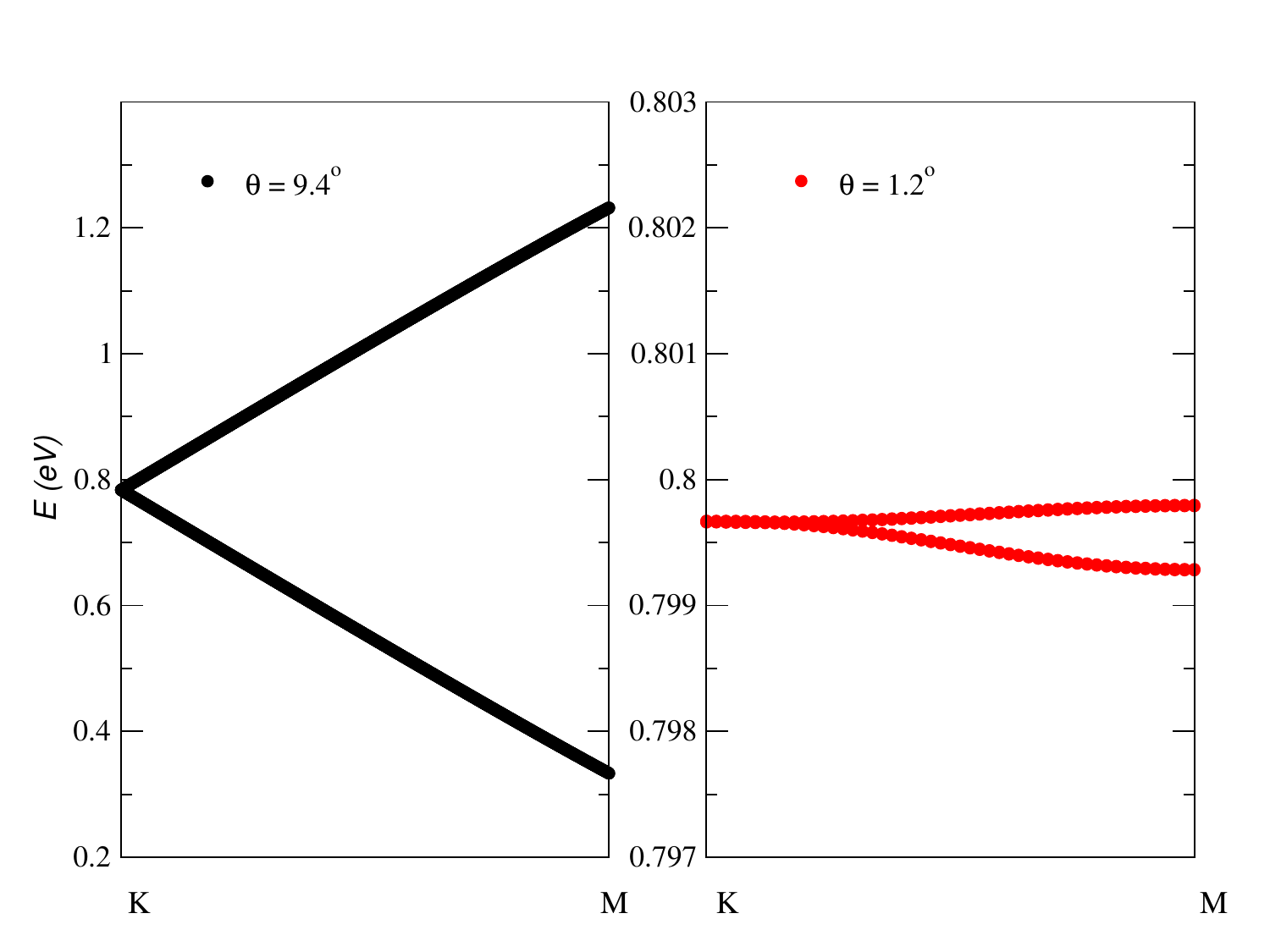}}
\caption{   Band structure around the charge neutrality point for bulk twisted bilayer graphene for twist angles $\theta = 9.4^{\circ}$ and $\theta = 1.2^{\circ}$). 
}
\label{bands} 
\end{figure*}

\subsection*{Twist angles beyond the magic angle}
\begin{figure*}[h]
\centering
\scalebox{0.9}{\includegraphics{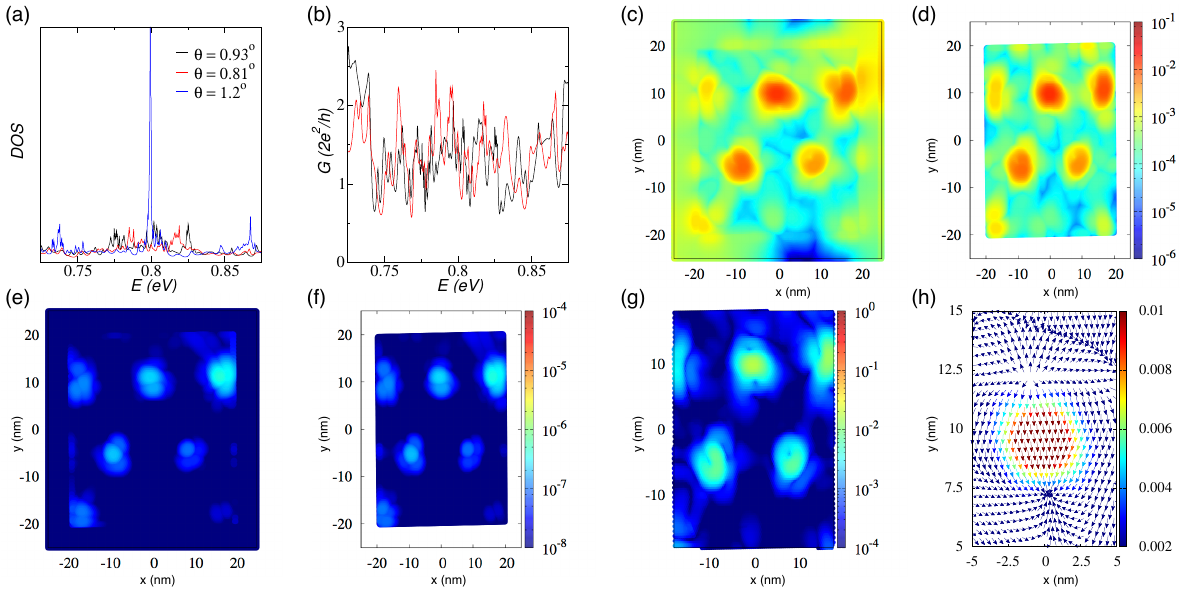}}
\caption{   DOS (a) and conductance (b) for twist angles $\theta = 0.93^{\circ} (i=35)$ and $\theta = 0.81^{\circ} (i=40)$. The blue line in panel (a) corresponds to the DOS for $\theta = 1.2^{\circ}$. Total current normalized by the total source-drain current per bond for bottom  (c) and top (d) layer. (e)-(f) In-plane magnetic moment calculated by $\vec m=\sum_{<ij>}I_{ij}(\vec{r}_i\times \vec{r}_j)/2$  in units of Bohr magneton for bottom and top layers. (g) Magnitude of the counterflow current ($|\vec{I}_m|$) normalized by the total source-drain current per bond. (h) Vectorial map of $\vec{I}_m$ over one AA-stacked region. In panels (c)-(h), the parameters are: $\theta = 0.81^{\circ}(i=40)$, $V_g = 0.1~\text{meV}$ and $V_{SD} = 100~\mu\text{V}$.
}
\label{figi40} 
\end{figure*}

For $\theta < 1^\circ$, there is no high DOS at the charge neutrality point as shown in  \Fref{figi40} (a).  Still, there is a high LDOS on the AA-stacked regions. For small twist angles, the Moir\'e periodicity $D = a/\sin(\theta/2) > 16.2~\text{nm}$ almost exceeds the dimension of the top layer and the few AA-stacked regions are not enough to produce a DOS peak at the CNP. From the transport point of view, the TBG efficiently scatters electrons into the available states producing interference as seen from the rapid oscillations in the conductance, see \Fref{figi40} (b). 

Regarding the main results presented in the main text, we continue observing high current density and in-plane magnetic moments on the AA-stacked regions since these effects are the result of having high LDOS on those regions. To assert the above mentioned, we plot for a TBG barrier with $\theta = 0.81^{\circ}(i=40)$, $V_g = 0.1~\text{meV}$ and $V_{SD} = 100~\mu\text{V}$ the magnitude of the electric current divided by the source-drain current per bond in \Fref{figi40}(c)-(d) and the in-plane magnetic moment in panels (e)-(f) calculated by the global formula $\vec m=\sum_{<ij>}I_{ij}(\vec{r}_i\times \vec{r}_j)/2$. The maps allow us to identify that in spite of the low number of AA-stacked regions the injected current still produces charge current and in-plane magnetic moments ``hot spots''. Moreover, because of the greater Moir\'e periodicity the in-plane magnetic moments appear totally localized on the central AA-stacked regions. The current counterflow maps (\Fref{figi40}(g)-(h)) also show  high values and preferred orientation on the same regions.

\section*{Local definition of the magnetic moment and chiral response}

\begin{figure*}[h]
\centering
\scalebox{0.9}{\includegraphics{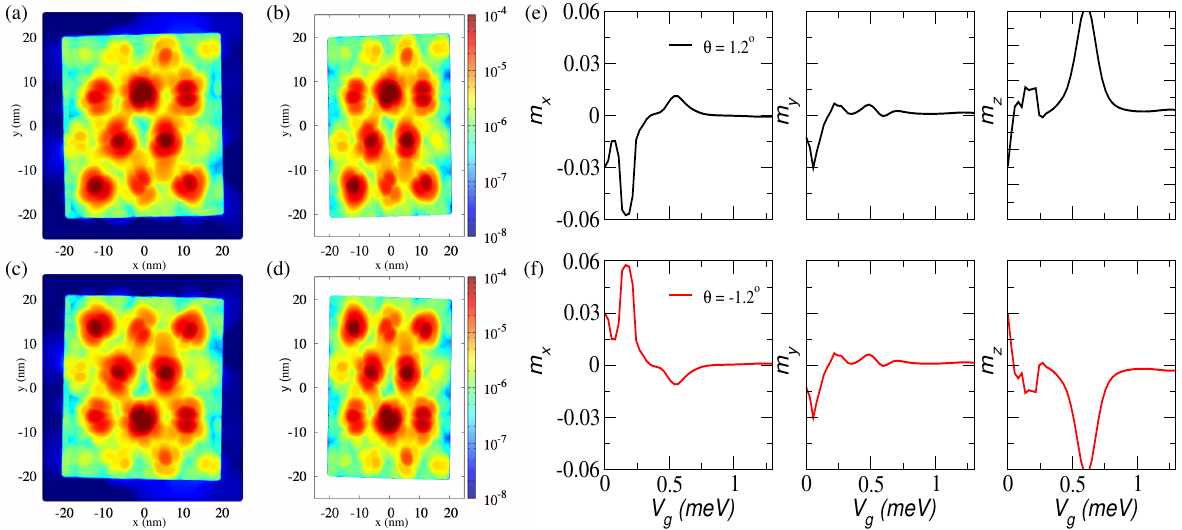}}
\caption{Map of the in-plane magnetic moment for  $\theta = 1.2^{\circ}$, see panels (a) and (b), and $\theta = - 1.2^{\circ}$, see panels (c) and (d), at gate voltage $V_g = 0.1~\text{meV}$. Components of the magnetic moment as function of $V_g$ for $\theta = 1.2^{\circ}$, see panel (e), and $\theta = -1.2^{\circ}$, see panel (f). In all panels, the magnetic moments are in  units of $\mu_B$ and $V_{SD} = 100~\mu \text{V}$
}
\label{MMmame} 
\end{figure*}

To check if the system size allows for a general analysis, we perform the calculations for a positive and negative twist angle. The infinite twisted bilayer system can be transformed from a positive to a negative twist angle by performing a parity-transformation $\vec{r}\to-\vec{r}$ and subsequent mirror-transformation ($\pi$ rotation around the $y$-axis). The position vector, current density, and magnetic moment transform accordingly, i.e., $(x,y,z)\to(x,-y,z)$, $(j_x,j_y,j_z)\to(j_x,-j_y,j_z)$, and $(m_x,m_y,m_z)\to(-m_x,m_y,-m_z)$.  

In \Fref{MMmame}, we can see that our finite system fullfil these requirements. Looking first at the map of in-plane magnetic moment ($V_g = 0.135~\text{meV}$ and $\theta = \pm 1.2^{\circ}$) in panels (a)-(d), a large magnetic moment is seen at the AA-stacked regions. These regions transforms as $(x,y,z)\to(x,-y,z)$ and can be linked to the high LDOS and current densities , present on the same spots. To underline the transformation of the magnetic moment, we plot the components $m_{x(y)(z)}$ in units of $\mu_B$ as function os $V_g$ in panels (e)-(f). Let us also mention that the sign change of $m_x$ under the transformation points at the chiral coupling of TBG as discussed in Ref. \citenum{Stauber18}.

\section*{Defining the in-plane magnetic moment}

\begin{figure*}[h]
\centering
\scalebox{1}{\includegraphics{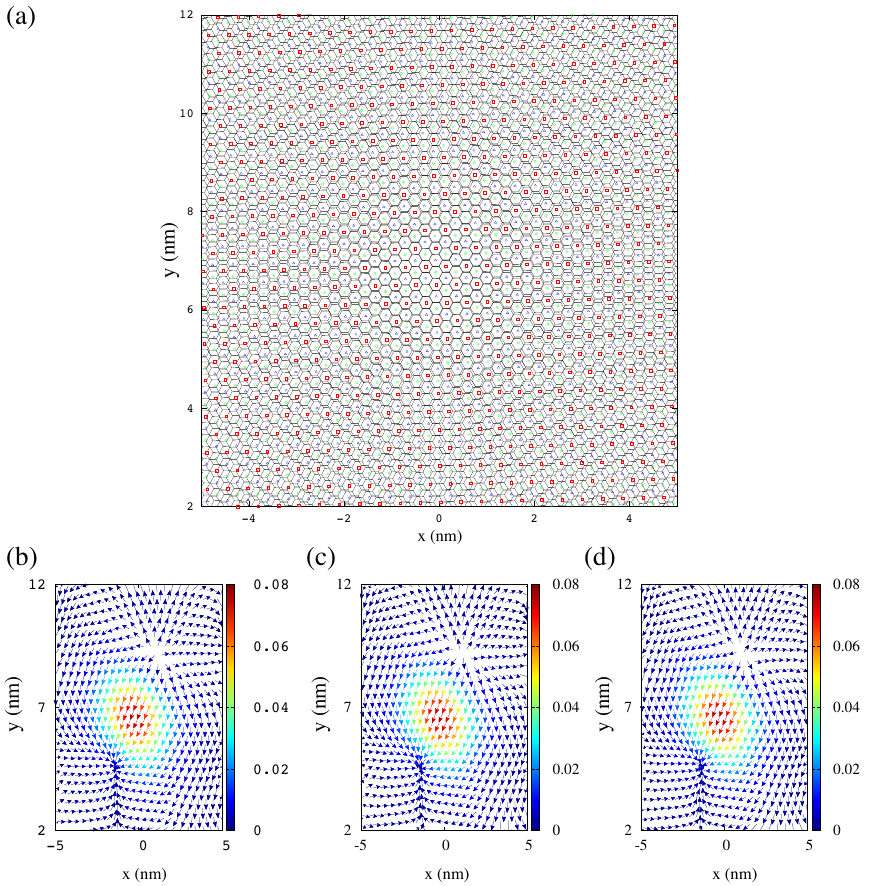}}
\caption{  (a) Real atomic lattice and the dual triangular lattices $\textcolor{red}{\square},~\textcolor{green}{\diamond}~\text{and}~\textcolor{blue}{\triangle}$ used to average the magnetic and total current. Vector map of $\vec{I}_m$ normalised by the total source-drain current per bond for triangular lattice (b) $\textcolor{red}{\square}$, (c) $\textcolor{green}{\diamond}$ and (d) $\textcolor{blue}{\triangle}$. In panel (b) to (d) $\theta = 1.2^{\circ}$, $V_g = 0.1~\text{meV}$ and $V_{SD} = 100~\mu\text{V}$
}
\label{imagavg} 
\end{figure*}

The calculation of the magnetic and total current at the atomic sites  is only well-defined in the AA-stacked regions where the atoms have approximately the same \textit{x} and \textit{y} coordinates. At these sites, $\vec{I}_{m}(x,y) = [\vec{I}_1(x,y)-\vec{I}_2(x,y)]/2$ and $\vec{I}_T(x,y) = \vec{I}_1(x,y)+\vec{I}_2(x,y)$ are well defined. To extend the calculation to other regions of the device, it is necessary to average the current on both layers. Our process is divided in two steps.

\begin{itemize}
\item The current on the top layer is averaged at the center of each hexagonal plaquette. To avoid double counting of the atomic sites, we average over the centers of every third hexagonal plaquette which form a triangular lattice with lattice parameter $3a$, where $a$ is the carbon-carbon distance. To cover up the top layer, we have three different possible triangular lattices. These are identified in \Fref{imagavg} (a) by the symbols \textcolor{red}{$\square$}, \textcolor{green}{$\diamond$} and \textcolor{blue}{$\triangle$}. We used these lattices to define the current for each in the hexagonal plaquettes of the top layer as:

\begin{equation}
\begin{split}
\vec{I}^{\textcolor{red}{\square}(\textcolor{green}{\diamond})(\textcolor{blue}{\triangle})}_2 =\sum_{s=1}^6\vec{I}_{2}(s)
\end{split}
\end{equation}

\item The current in the bottom layer, $\vec{I}_1$, is averaged using the same triangular lattice defined for the top layer, but this time we select the atomic sites  within a radius $R=1.5a$: 
\begin{equation}
\begin{split}
\vec{I}^{\textcolor{red}{\square}(\textcolor{green}{\diamond})(\textcolor{blue}{\triangle})}_1 = \sum_{<1.5a}\vec{I}_{1}\;.
\end{split}
\end{equation}
\end{itemize}
Using the above procedure the coordinates of the top and bottom current are the same and we can proceed to calculate $\vec{I}_{m}$ and $\vec{I}_{T}$. 

To confirm that the results obtained do not (strongly) depend on  triangular lattice used, we present the resulting magnetic current using the local definition in \Fref{imagavg} (b) - (d). It is clearly appreciated  that the enhanced counterflow current in  AA-stacked regions remains a robust feature irrespective of details of the calculation method.

\section*{Perturbations}

\subsection*{Lattice relaxation}
Let us analyze in more detail the effect of lattice relaxation following the approach by Nam and Koshino.\cite{Nam17,NamErratum} We first present the source to drain current for the relaxed lattice that we used to normalise the current maps in the main text. In \Fref{relaxall} (a), we show the calculated current as function of the gate voltage for $\theta = 1.2^{\circ}$, $V_{SD}=100~\mu\text{eV}$.

\begin{figure*}[h]
\centering
\scalebox{1}{\includegraphics{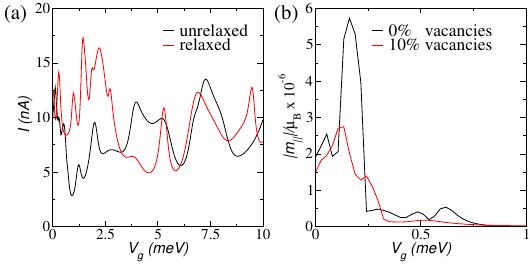}}
\caption{  (a) Source-drain current for the (un)relaxed device.  The relaxation parameters were taken from Nam and Koshino.\cite{Nam17,NamErratum} (b) In-plane magnetic moment per site as function of the gate voltage around the AA-stacked region between $-5\leq x/\text{nm} \leq 5$ and $2\leq y/\text{nm} \leq 12$ calculated via the global  formula of the magnetic moment. The values were obtained by averaging over 5 randomly distributed vacancy realizations.
 In all panels, we considered  $\theta = 1.2^{\circ}$ and $V_{SD}=100~\mu\text{eV}$.
}
\label{relaxall} 
\end{figure*}

\subsection*{Vacancies}
Vacancies in graphene induce the formation of localised states that can perturb the current distribution.\cite{PhysRevB.82.165438} In our device with armchair edges, we observe a reduction in the value of the in-plane magnetic moment for the region between $-5\leq x/\text{nm} \leq 5$ and $2\leq y/\text{nm} \leq 12$. This is shown in Fig. \ref{relaxall} (b) for a density of 10\% of vacancies in the top flake edges having considered 5 randomly distributed vacancy realizations. We observe that the overall in-plane magnetic moment is robust against vacancies

\subsection*{Zigzag edges}
\begin{figure*}[h]
\centering
\scalebox{0.9}{\includegraphics{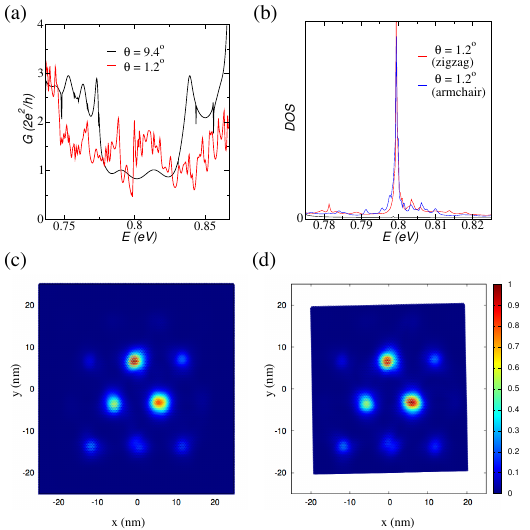}}
\caption{For the  TBG barrier with the zigzag graphene nanoribbon as the bottom layer. (a) Conductance. (b) DOS. The lower panels show the local Density of States for $\theta = 1.2^{\circ} (i=27)$ and $E_F= 0.8~\text{eV}$: (c) bottom layer; (d) top layer.
}
\label{fig:zigzag} 
\end{figure*}

We finally present the results for the TBG barrier embedded on top of a zigzag nanoribbon. For large angles, both layers are decoupled and the conductance is the same as the conductance of the single monolayer with zigzag edges. However, compared with armchair case the Fabry-Perot oscillation are more pronounced, see \Fref{fig:zigzag} (a). Similar to the armchair case, we observe a high DOS at the CNP. Also, the local DOS at this energy shows wavefunction localisation on the AA-stacked regions.

\section*{Spatial Distribution of Currents in Twisted Bilayer Graphene in the Continuum Model}
\begin{figure}
\centering
\includegraphics[width=0.5\columnwidth]{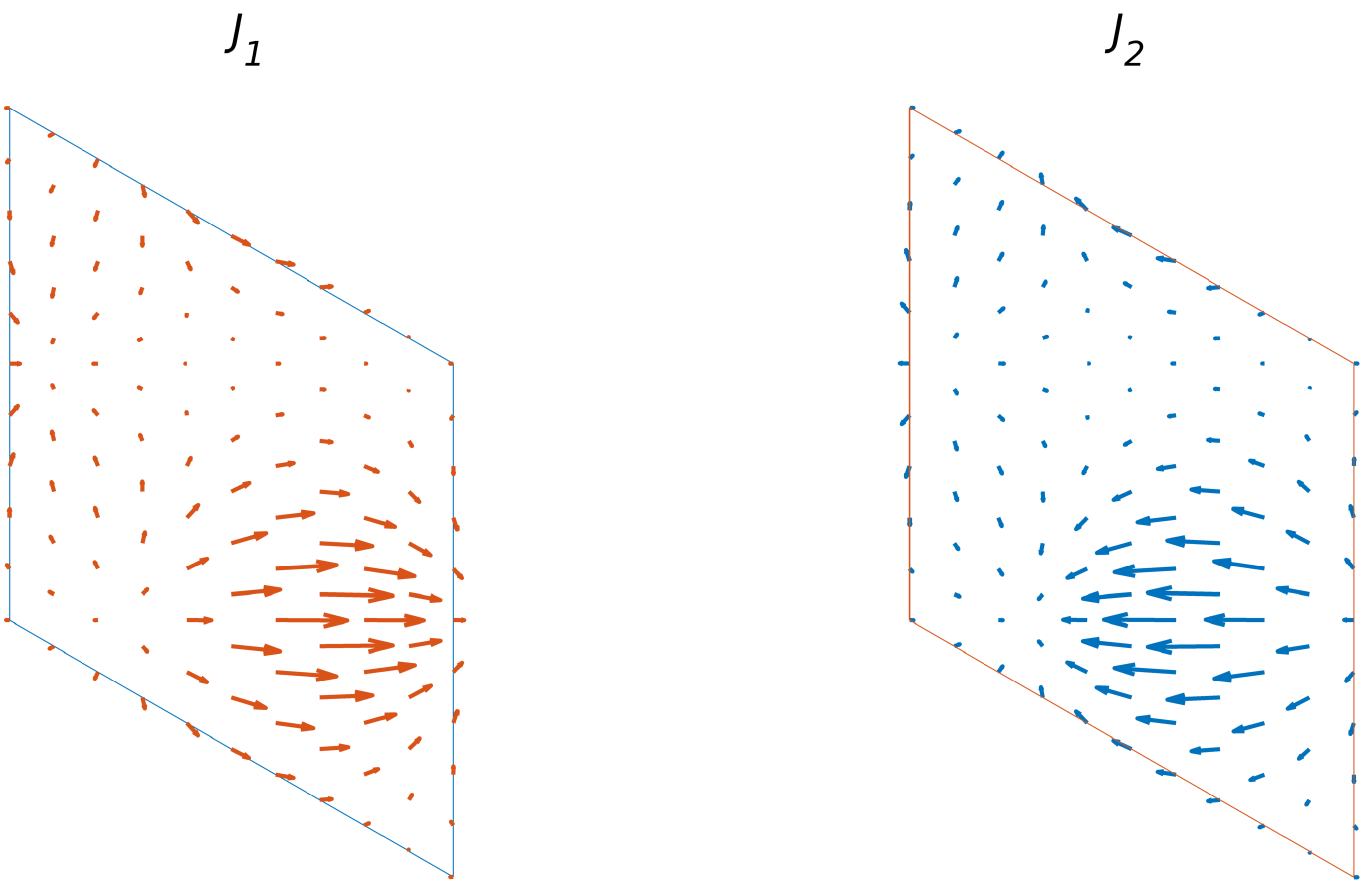}
\caption{ :  Current map (band average) within the Moir\'e unit cell for twist angle $\theta=1.05^{\circ} (i=31)$, as obtained from the continuum model of Ref. \cite{Lopes07}. The current is the response to the adiabatic introduction of a uniform vector potential along the negative $x$ axis acting {\em only}  on the layer 2.  The current is strongly enhanced around the AA-stacked region (center of lower triangle) and minimal around the AB-stacked (corners) and BA-stacked (center of upper triangle) regions. Notice that the current of layer 2 is opposite to the field direction giving rise to a paramagnetic response. 
\label{J12}}
\end{figure}
\begin{figure}
\centering
\includegraphics[width=0.5\columnwidth]{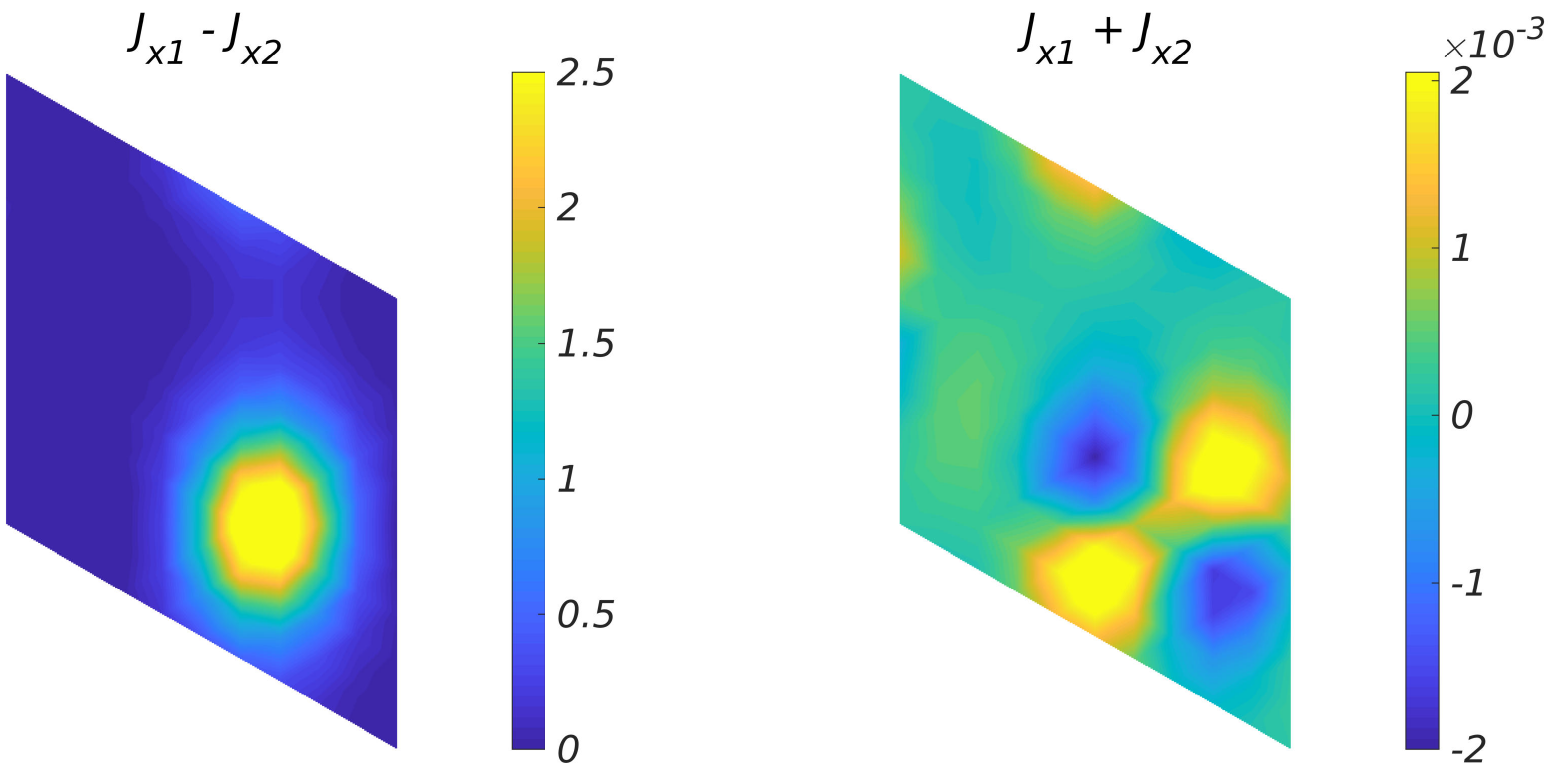}
\caption{ :  Left panel: Counterflow current map (band average) within the  Moir\'e unit cell. Right panel: Total current map (band average) within the  Moir\'e unit cell. Units are arbitrary but the same for both panels and notice the huge difference in scales between both cases.
\label{Jminus}}
\end{figure}
\begin{figure}
\centering
\includegraphics[width=0.5\columnwidth]{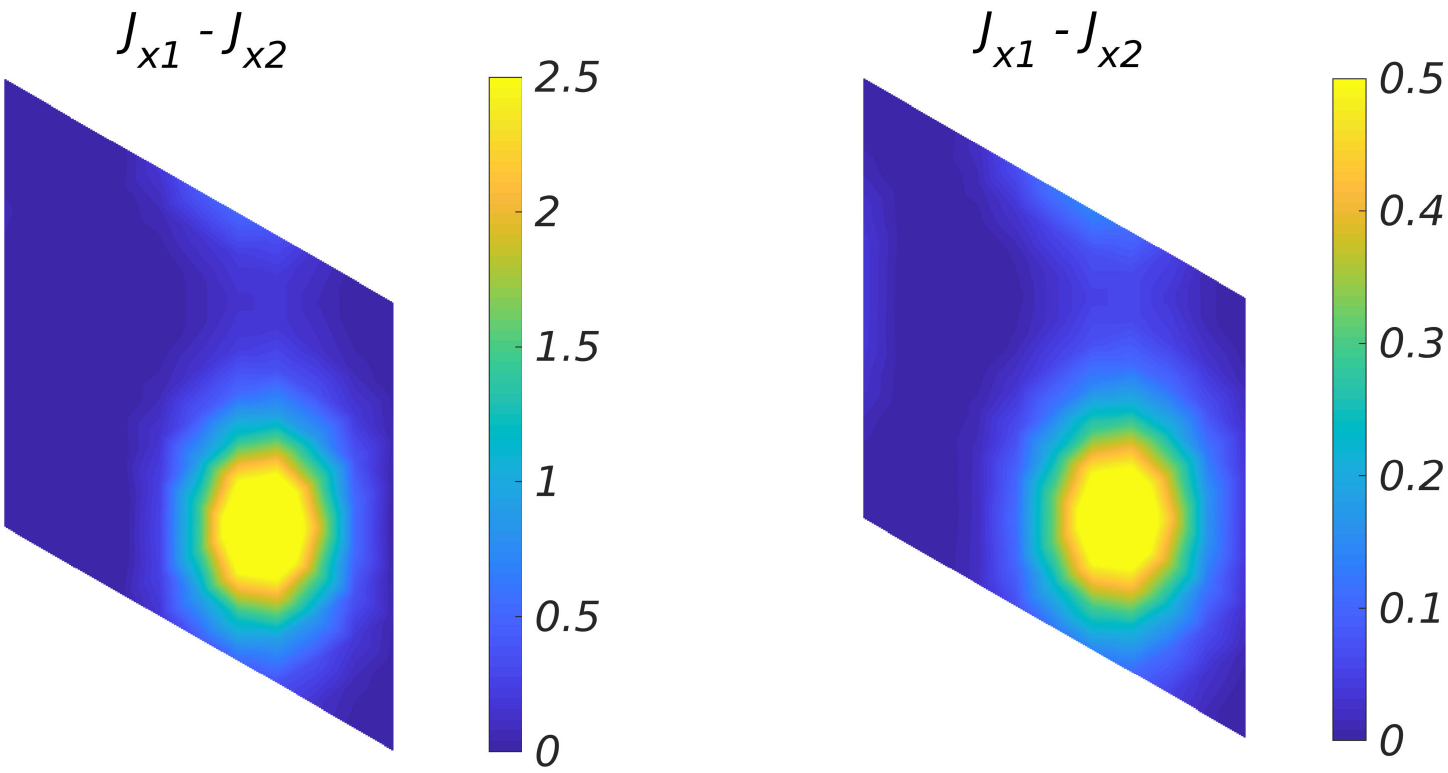}
\caption{ :   Left panel: Counterflow current (band average) for the unrelaxed lattice within the  Moir\'e unit cell. Right Panel: As in the left panel for the lattice with relaxation. 
\label{pcolor}}
\end{figure}

Here we discuss the spatial distribution of currents induced by the adiabatic introduction of a uniform vector potential along the negative $x$ axis acting only  on the layer 2, within the continuum model of Lopes dos Santos et al.\cite{Lopes07} The transient electric field points towards the positive $x$ axis and is therefore restricted to the layer 2, but currents are generated in both layers. We expect this asymmetric driving to best mimic the scattering calculation of the main text, even though the geometry differs: here the calculation corresponds to an infinite system. The twist angle is $\theta=1.05^{\circ} (i=31)$ and the intra and interlayer hopping parameters  
are given by $t=3$eV and $t_{\perp}=0.12$eV (the value quoted for $t_{\perp}$ in the SI of Ref. \cite{Stauber18} should be divided by $3$). The calculation is  standard linear response for the continuum model\cite{Stauber18}, adapted to obtain the response current at position $\bm r$, given by
\begin{equation}
\bm j  (\bm r) = \frac{e v_F}{2} (|{\bm r}\rangle\langle{\bm r}| \bm \sigma + \bm \sigma|{\bm r}\rangle\langle{\bm r}| )
,\end{equation}
where $\bm \sigma$ are pseudospin (current) operators. The calculation is restricted to Fermi levels within the lowest electron and hole bands around the neutrality point. Main results are:

1) The current is largest in the AA-stacked region, and opposite in both layers with near cancellation, as expected from previous work\cite{Stauber18,Stauber18b}. This is a generic property of the considered bands, as shown in \Fref{J12}, where the currents averaged for Fermi levels spanning the lowest electron and hole bands is presented.
 
2) The near cancellation makes the counterflow  ({\em magnetic}) current, $ \bm J_1 -\bm J_2 $, to be largely enhanced  in the AA-stacked regions as compared to the total current, $ \bm J_1 +\bm J_2 $. \ This is shown in \Fref{Jminus} where the counterflow current exceeds  the total current by three orders of magnitude. In fact, this is a conservative estimate  because \Fref{Jminus} represents the band average whereas the enhancement factor can be  nominally infinite at the Dirac point, where the total current should vanish but the counterfow does not\cite{Stauber18,Stauber18b}.

3) We have also mimicked the presence of lattice relaxation by a $20$-percent  weakening of the AA interlayer hopping as compared to the AB one in the continuum model. The previous conclusions are hardly affected by this change, as shown in \Fref{pcolor} where the counterflow currents are represented for both the undistorted and distorted cases. 

All these features  agree with the main message of this work: enhanced counterflow in AA-stacked regions close to the magic angle. It is interesting to remark that, although the total current flows in the positive $x$-direction, which coincides with the (transient) electric field as expected, the current in the layer  where the field is applied (layer 2) runs opposite to the field, see right panel of \Fref{J12}. This fact is at the heart of  the large paramagnetic response previously reported.\cite{Stauber18,Stauber18b} 
\end{widetext}

\bibliographystyle{unsrt}
\bibliography{TBGbibV1}

\end{document}